\documentclass[
    ,final            
  ]
  {aipproc}

\layoutstyle{6x9}

\def\gtsim {>\kern-1.2em\lower1.1ex\hbox{$\sim$}~}   
\def\ltsim {<\kern-1.2em\lower1.1ex\hbox{$\sim$}~}   
\def \apj {ApJ}

\def \mnras {MNRAS}

\begin{document}

\title{Chemodynamical simulations with variable IMF}


\author{Chiaki Kobayashi}{
address={The Australian National University,
Mt. Stromlo Observatory, Cotter Rd., Weston ACT 2611}
}

\begin{abstract}
Using self-consistent chemodynamical simulations including star formation, supernova feedback, and chemical enrichment, I show the dependence of cosmic star formation and chemical enrichment histories on the initial mass function (IMF).
The effects of Pop-III IMF can be only seen in the elemental abundance ratios at $z\gtsim 4$ or [Fe/H] $\ltsim -2$.
The preferable IMF has a flatter slope in the case of high star formation rate (SFR) and smaller upper-mass ($\sim 20M_\odot$) in the case of low SFR, which is consistent with the observed elemental abundances of dwarf spheroidal galaxies. However, the [$\alpha$/Fe] problem of elliptical galaxies may require other solutions.
\end{abstract}

\maketitle


\section{Introduction}
\vspace*{-2mm}

While the evolution of the dark matter
is reasonably well understood, the evolution of the baryonic component is much
less certain because of the complexity of the relevant physical processes, such
as star formation and feedback.
With the commonly employed,
schematic star formation criteria alone, the predicted star formation rates (SFRs)
are higher than what is compatible with the observed luminosity density. 
With supernova and hypernova feedback, the cosmic star formation and metal enrichment histories are well reproduced, as well as the mass-metallicity relation of galaxies \cite{kob07}.

In the chemodynamical simulations, the initial mass functions (IMF) is one of the most important assumptions.
In the scheme introduced by \cite{kob04}, because of the resolutions, one star particle is treated as a single stellar population that have the coeval age and chemical composition, and the mass of individual stars are distributed according to the IMF.
The amount of supernova feedback and chemical enrichment from the star particle to the interstellar medium depends on the IMF.
Supernovae inject not only thermal energy
but also heavy elements into the interstellar medium,
which can enhance star formation.
Chemical enrichment must be solved as well as energy feedback.
Different supernovae produce different heavy elements with different timescales.
Therefore, chemical abundances, namely, elemental abundance ratios can put a constraint on the IMF from the detailed comparison between simulations and observations.
In this paper, we show the effects of variable IMF in our cosmological simulations. The details of the hydrodynamic code are described in \cite{kob07}, and the nucleosynthesis yields of Type II Supernovae (SNe II) are in \cite{kob06}, and the progenitor model of Type Ia Supernovae (SNe Ia) is in \cite{kob09}.

\begin{figure}
  \includegraphics[height=.25\textheight]{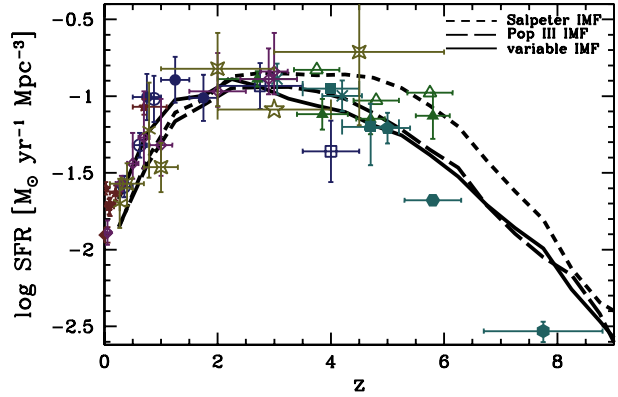}
  \caption{Cosmic star formations histories with the Salpeter IMF (short-dashed line), the Pop-III IMF (long-dashed line), and the variable IMF (solid line). See \cite{kob07} for the observational data sources (dots).}
\end{figure}


The IMF is shown by Salpeter (1955) from the observed luminosity functions of stars in the solar neighborhood, where the number of stars with a given mass is given by the power-law with a slope of $-1.35$.
Recently, \cite{kro07} showed a similar slope ($-1.3\pm0.5$) for $m \ge 0.5 M_\odot$, which is flatter than in the Millar-Scalo (1979)'s and Scalo (1986)'s IMFs.
For $m < 0.5 M_\odot$, the slope is shallower and the number of brown dwarfs are smaller than in the Salpeter IMF.
In chemical evolution models, however, metal enrichment is proceeded only by stars with $m \ge 0.5 M_\odot$.
If a suitable value is chosen for the lower mass-limit of the IMF, the Salpeter IMF can give equal results with the Kroupa IMF \cite{kob09}.
Therefore, we adopt the single slope for a mass range of $m=0.07-120 M_\odot$.
The lower mass-limit of $M_\ell=0.07 M_\odot$ is chosen for the adopted nucleosynthesis yields to reproduce the observed metallicity distribution function in the solar neighborhood \cite{kob06}.
This does not necessarily mean that there is no star at $m < 0.07 M_\odot$. The requirement is that the mass fraction of $m \ge 0.5 M_\odot$ is 45\%.

From the theory of primordial star formation, the mass accretion does not stop and the fragmentation does not occur, and thus the first star should be very massive. The stars with $\gtsim 130 M_\odot$ explode as Pair Instability supernovae, but the contribution should be very small since the elemental abundance pattern should be very much different and no such signature has been observed \cite{kob06}.
Therefore, in the Population III (Pop-III) stars, we set the lower mass of $M_\ell=20M_\odot$, the upper mass of $M_{\rm u}=120M_\odot$, and the slope of $x=0$.
From the observations of stellar populations in early-type galaxies, the IMF slope seems to be as flat as $x=-1.1$ (e.g., \cite{hop08}). Thus we assume the flatter IMF for the high SFR mode.
From the observations of elemental abundances in dwarf Spheroidal galaxies (namely, [Mn/Fe]), there seems not to be the contribution of massive supernovae and hypernovae (Kobayashi 2010, in preparation). Thus we assume the upper mass of $M_{\rm u}=20M_\odot$ for the low SFR mode.
This is reasonable since the number of massive stars should be very small in the case of low SFRs.

\vspace*{-2mm}
\section{Results}
\vspace*{-2mm}

Figure 1 shows the cosmic star formations histories with i) the Salpeter IMF, ii) the flatter IMF for Pop-III stars with $x=0$, and iii) the Pop-III IMF and the variable IMF with $x=-1.1$ for high SFR and $M_{\rm u}=20M_\odot$ for low SFR.
The model with Salpeter IMF 
is the same as in \cite{kob07}, but 
with the WMAP-5 cosmology.
With the Pop-III IMF, 
the SFR is reduced by a factor of $\sim 5$ at $z \gtsim 4$ due to the larger feedback, but is not affected at $z \ltsim 2$ since only negligible number of primordial stars form at later epoch.
With the variable IMF, 
the global star formation history is not affected very much.
The SFR at $z\sim4$ is slightly smaller 
because most star formation takes place at the high SFR mode, which is set to have stronger feedback with the flatter IMF.
The SFR at $z \ltsim 2$ is slightly larger 
due to the smaller feedback of the low SFR mode.

\begin{figure}
  \includegraphics[height=.3\textheight]{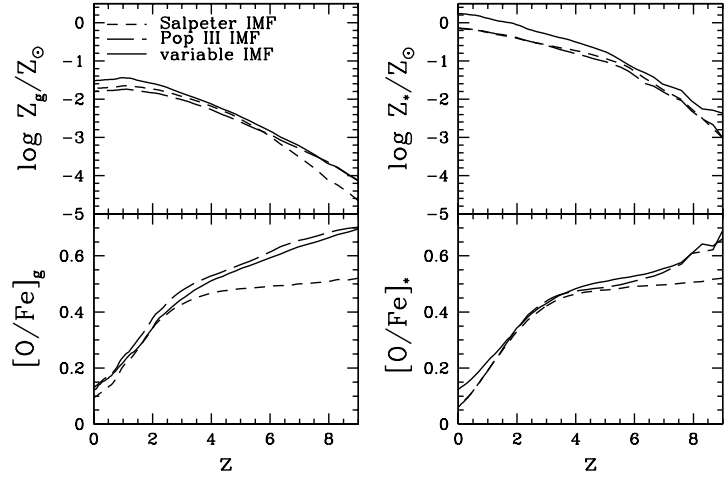}
  \caption{Cosmic chemical enrichment histories with the Salpeter IMF (short-dashed line), the Pop-III IMF (long-dashed line), and the variable IMF (solid line).}
\end{figure}

Figure 2 shows the redshift evolution of metallicity (upper panels) and elemental abundance ratios (lower panels), mass-weighted in gas phase (right panels) and luminosity weighted in stellar populations (left panels). Here the oxygen to iron ratios are shown, but other $\alpha$ elements (O, Mg, Si, S, and Ca) show the same trends.
As stars form and die, metallicity increases.
At the early stage, only SNe II contribute, and [O/Fe] stays constant until $z\sim4$ (short-dashed lines). After $0.1$ Gyr from the formation of stars with [Fe/H] $\sim -1$, which depends on the adopted SN Ia progenitor model, SNe Ia produce more Fe and O.
The decrease of [O/Fe] from $z\sim3$ to $z=0$ is caused by the delayed enrichment of SNe Ia.

With the Pop-III IMF (long-dashed lines), because the net yields are much larger, metallicity is increased by $0.5$ dex at $z \gtsim 6$, but is not affected at $z \ltsim 6$.
Among SNe II, massive stars produce more oxygen, and thus [O/Fe] is as large as 0.7 at $z\sim8$.
At later time, less-massive SNe II of normal IMF contribute, which causes the decrease from $z\sim8$ to $z\sim4$.
However, this trend is not seen so much in the stellar populations.

With the variable IMF (solid lines), the global chemical evolution depends mainly on the high SFR mode. The IMF slope is set to be flatter, the net yields are larger, and the net [O/Fe] is also larger because more massive stars produce more oxygen.
At $z=0$, metallicity is larger by $0.4$ dex, and the present stellar metallicity is $\sim 3Z_\odot$, which may be too large compared with observations.
The stellar [O/Fe] is as large as $0.1$, which may be still smaller than observations.

\begin{figure}
  \includegraphics[height=.345\textheight]{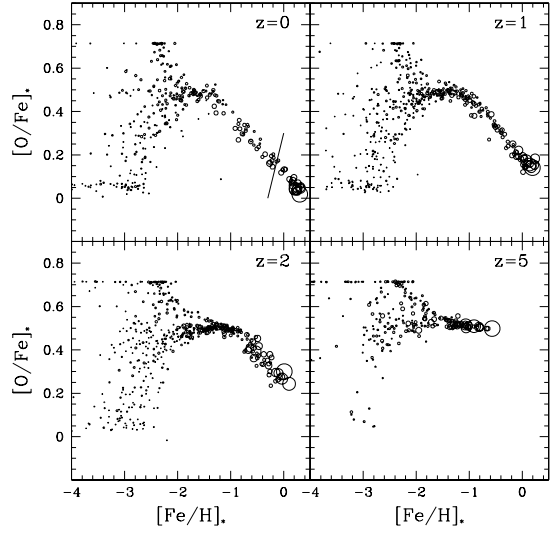}
  \caption{Redshift evolution of the [O/Fe]-[Fe/H] relation of luminosity weighted stellar populations in galaxies (dots) with the variable IMF. The solid line is for the observed relation \cite{spr10}.}
\end{figure}

Figure 3 shows the [O/Fe]-[Fe/H] relations of luminosity weighted stellar populations in galaxies with the variable IMF. The size of dots represents the stellar mass of galaxies; more massive galaxies have larger metallicity. In other words, the observed mass-metallicity relations are well reproduced with the simulated galaxies because of the mass-dependent galactic winds \cite{kob07}.

In the CDM Universe, the duration of star formation tends to be longer and the SN Ia contribution is larger for metal-rich massive galaxies. Therefore, [O/Fe] tends to be smaller for massive galaxies, which is inconsistent with the observational estimates; the solid line shows the observed relation by \cite{spr10}, where the global [$\alpha$/Fe] of the whole galaxies is estimated from abundance radial gradients.
The flatter IMF cannot provide enough high [$\alpha$/Fe] in the updated nucleosynthesis yields \cite{kob06}. Some additional physical processes may be required to solve this problem.

The characteristic of the variable IMF in this paper is seen at [Fe/H] $\sim 2$, where [O/Fe] spans in a range of $0-0.7$, which is consistent with the observed [$\alpha$/Fe] of dwarf Spheroidal galaxies.
The maximum [O/Fe] $=0.7$ is determined by the net [O/Fe] of the Pop-III IMF. The low [O/Fe] is caused not by the SN Ia contribution because the metallicity is too small to cause SNe Ia in our SN Ia model, but by the smaller upper mass of $M_{\rm u}=20M_\odot$ of the IMF in the low SFR mode.
In metal-poor dwarf galaxies, different from Galactic halo, [$\alpha$/Fe] tends to be $\ltsim 0.5$ and [Mn/Fe] $\ltsim 0$.

\vspace*{-2.5mm}

\bibliographystyle{aipproc}   

\end{document}